\documentclass[a4paper]{article}
\usepackage[utf8]{inputenc}
\usepackage[english]{babel}
\usepackage{amsfonts}
\usepackage{amsmath}
\usepackage{caption}
\usepackage[pdftex]{graphicx}
\usepackage{url}
\usepackage{hyperref}
\usepackage{placeins}
\usepackage{authblk}
\hypersetup{colorlinks=true}

\graphicspath{ {./figures/} }

\usepackage{amsthm}
\theoremstyle{definition}

\usepackage{csquotes}
\usepackage{listings}
\usepackage{xcolor}

\definecolor{codegreen}{rgb}{0,0.6,0}
\definecolor{codegray}{rgb}{0.5,0.5,0.5}
\definecolor{codepurple}{rgb}{0.58,0,0.82}
\definecolor{backcolour}{rgb}{0.95,0.95,0.92}

\lstdefinestyle{mystyle}{
  backgroundcolor=\color{backcolour},   commentstyle=\color{codegreen},
  keywordstyle=\color{magenta},
  numberstyle=\tiny\color{codegray},
  stringstyle=\color{codepurple},
  basicstyle=\ttfamily\footnotesize,
  breakatwhitespace=false,         
  breaklines=true,                 
  captionpos=b,                    
  keepspaces=true,                 
  numbers=left,                    
  numbersep=5pt,                  
  showspaces=false,                
  showstringspaces=false,
  showtabs=false,                  
  tabsize=2
}

\usepackage{biblatex}
\begin{document}


\title{Constructing trading strategy ensembles by \newline classifying market states}

\author[1]{Michal Balcerak\thanks{m1balcerak@gmail.com, balcerak.michal@stud.uni-goettingen.de }}
\author[2]{Thomas Schmelzer\thanks{thomas.schmelzer@gmail.com, thomas.schmelzer@unil.ch}}
\affil[1]{Institute for Theoretical Physics,\newline 
Georg-August-Universität Göttingen, Germany}

\affil[2]{Faculty of Business and Economics (HEC Lausanne),\newline University of Lausanne, Switzerland}

\renewcommand\Authands{ and }
\date{\vspace{-4ex}}

\maketitle

Rather than directly predicting future prices or returns, we follow a more recent trend in asset management and classify the state of a market based on labels. We use numerous standard labels and even construct our own ones. 

The labels rely on future data to be calculated, and can be used a target for training a market state classifier using an appropriate set of market features, e.g. moving averages. The construction of those features relies on their \textit{label separation power}. Only a set of reasonable distinct features can approximate the labels.

For each label we use a specific neural network to classify the state using the market features from our \textit{feature space}.
Each classifier gives a probability to buy or to sell and combining all their recommendations (here only done in a linear way) results in what we call a \textit{trading strategy}.
There are many such strategies and some of them are somewhat dubious and misleading. We construct our own metric based on past returns but penalizing for a low number of transactions or small capital involvement. Only top score-performance-wise trading strategies end up in final ensembles.

Using the Bitcoin market we show that the strategy ensembles outperform both in returns and risk-adjusted returns in the out-of-sample period. Even more so we demonstrate that there is a clear correlation between the success achieved in the \textit{past} (if measured in our custom metric) and the \textit{future}.


\tableofcontents
\section{Introduction}

Using neural networks to predict financial time series data is today widely regarded as the old unfulfilled dream of quantitative finance. An idea would be to apply supervised learning and train a neural network with sub-windows of a time series to predict the next data point(s).
So instead of using images of dogs and cats we use at some time $t$ the last $n$ points of a time series to predict a point following at some time $t' > t$. Given the non-stationary nature of time series market data and low signal-to-noise ratios, this is a rather ambitious problem.

For instance, rather than using $n$ prices (or returns), we reduce the dimensionality of the problem by using $m << n$ features based on the very same $n$ points, i.e.\@ an optimal combination of $m$ moving averages. Such questions have typically been addressed by linear regression. However, linear regression fails to exploit any non-linear effects between the features.

We do not stop by only modifying the input - we also alter the goals of our predictions. Rather than aiming for a (noisy) price trajectory we ask simpler questions more suitable for the machinery of machine learning. Our goal is to quantify the probability $p$ of a market being in a class or category $c$ or moving into one within the next hours or minutes. This could be the probability for a trend reversion or a spike in volatility or volume. We rely on labels as recently made popular by López de Prado \cite{Lopez1} but also create some on our own. The flexibility of labels allows us to design a strategy by emphasizing effects we try to cover.

For each label we ask for an optimal set of $m$ features to approximate them.
These features, through a classifier, induce a probability for the market to be in a particular label-class.
We then ask for an optimal linear combination of those probabilities to execute trades. Rather than looking at a Sharpe ratio in an out-of-sample period we construct robust variations of this concept and penalize for a lack of trading activity, etc. Although we don't aim directly for it we observe high Sharpe ratios and attractive returns as an unavoidable collateral side effect.

\section{Labeling}
\label{sec:labeling}

We describe a market by a time series of datapoints $p_{t_0}, p_{t_1}, \ldots$. Predicting unseen price data is a hard problem often resulting in the notorious estimate that the next price is just the last observed price.

Rather than aiming for the next price, we argue that the market is currently in a particular label-class which we ultimately want to identify without using any unseen future data. 

Throughout this work we distinguish three such label-classes and identify them with the actions we intend to take:

\begin{itemize}
    \item Buy. The market may start or continue to rise over the next few periods
    \item Sell. The market may drop over the next few periods, the volume may drop significantly or there is a spike in volatility. 
    \item Neutral. We do nothing.
\end{itemize}
Obviously identifying buy opportunities is trivial with a good level of hindsight. Looking at a historic time series we can identify numerous buy opportunities. This process is subject to some constraints we set, e.g.\ we may argue that the price at time $t$ was a buy opportunity if we see a significant rise over the next minutes following $t$. 

We call the process of classifying over time the \emph{labeling} of a time series.
So the particular \emph{label} is a time series mapping $p_t$ to one of the three classes.

Numerous such labels can and should be used. We use the popular \emph{threshold label} to discuss the concept.
We define the return over the period $t_i$ to $t_{i+1}$ as
\[
r_{i, i+1} = \frac{p_{t_{i+1}}}{p_{t_i}} - 1.
\]
We introduce a threshold $\tau$ and use

\begin{equation}
        y_i=
        \left\{ \begin{array}{ll}
            sgn(r_{i,i+1}) & |r_{i, i+1}| > \tau \\
            0 & \text{otherwise}
        \end{array} \right. 
\end{equation}

Note that $y_i = 0$ if $|r_{i, i+1}| \leq \tau$ otherwise $y_i = 1$ or $y_i=-1$.
 
We also use a continuous companion of this label function. We use a continuous interpolation between the two labels $1$ and $-1$. So if $|r_{i, i+1}| <= \tau$ we use instead of $y_i = 0$ the function
\[
y_i =  \left(\frac{r_{i, i+1}}{\tau}\right)^3. 
\]
We could use a simpler linear term. However, in our experiments we have made better experiences with this particular choice.

So given a historic time series with all its price jumps and chaotic behaviour we reduce it to a time series just oscillating between three label-classes. Obviously we loose some information in this process but one could also argue we  emphasize the information we really care about. And we can always combine multiple labels.

Identifying the moments we have missed to make a profit can help to evaluate the quality of a strategy, however, its inherent delay renders it of limited use in a live trading setup.

The idea is to approximate the labels with market features (i.e.\@ technical trading indicators) that do not use any future data.
Once in live trading, we can live update the indicators and therefore talk about label-classes predictions.
The threshold is often made dynamic using estimates for the current volatility. 

We use a variation of this idea where rather than $p_{t_{i+1}}$ in the definition of the return we use a moving average of prices following $t_i$.

The construction of such labels is an exercise only done during the training phase of the strategies. 
Running a backtest based on the actions induced by the labels over this training period would be a severe mistake.

Although it would be possible to have labels based on all sorts of financial data, e.g. volume, we use here exclusively labels based on price data.

\section{Market representation}
The central idea of this paper is to approximate the labels with a set of functions, referred to also as market features. The functions we use are standard technical indicators.
The art is to resolve the labels in a small set of such parametrized functions. Those parameters are chosen in a way to maximize the \emph{label separation power} of those functions.

Although the arsenal of orthogonal functions, i.e.\@ a set of sin waves, is generally a great choice for approximations, we believe it is not suitable to capture market dynamics. A Fourier transform of the label would learn everything about the seasonality of this label but is of very limited generalization in an out-of-sample period.

We present our ideas using a toy example of only two functions with one free parameter each. In  \nameref{subs:appendix_2} we give a complete list of features we have used.

The set of feature functions we identify as \emph{feature space}.
The parameters are not completely free. They are integer numbers from intervals we define. Hence we can pick for each label from a finite set of such features.
\label{sec:market_representation}

\subsection{Feature space}
To illustrate a \textit{feature space} on an relatively simple example, let us define it as 2 indicators with some possible parameters: \\
\textbf{Example feature space:}
\begin{itemize}
    \item feature 1: $A[X]$, VWAP - SMA($X$) where $X \in [2,10]$ [minutes]
    \item feature 2: $B[Y]$, VWAP - SMA($Y$) where $Y \in [30,60]$ [minutes]
\end{itemize}
where VWAP stands for volume weighted average price of a givin minute and SMA($Z$) stands for a moving average of the last $Z$ minutes.

This set of two features has one feature that looks at relatively short term time horizon and one feature with relatively long term. 
We normalise their values to $[-1,1]$ using local scaling by standard deviation and arctangent function.

Let us define a label as one of the threshold labels: 1.5 \% price change in 5 minute window. We now face a dilemma - which features from the feature space should we use? There are 279 candidates (9 different feature 1 and 31 different feature 2).

Let us fix parameters to acquire 2 possible feature sets from the feature space and solve the dilemma there: 

\begin{itemize}
    \item feature set 1:\begin{equation}\label{eqn:featuresA} A[5], B[50]  \end{equation}
    \item feature set 2:\begin{equation}\label{eqn:featuresB} A[10], B[30]  \end{equation}
\end{itemize}

\begin{figure}[!h]
    \centering
    \textbf{BTC/USDT market and its representation}
    \includegraphics[width=\textwidth]{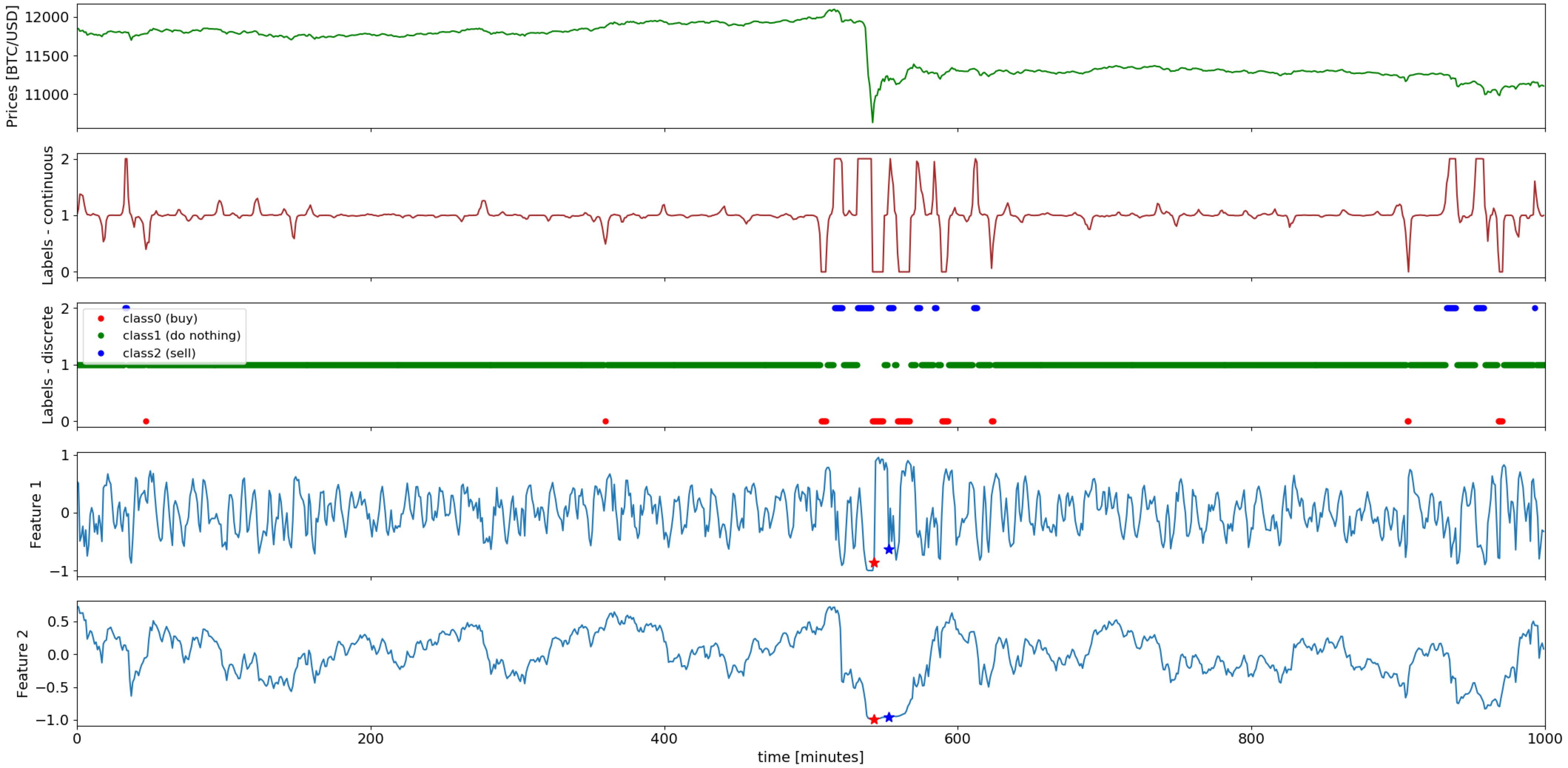}
    \caption{Market VWMP, threshold label with 1.2\% price change, 5 minutes time horizon and its Eq. \ref{eqn:featuresA} market representation. Zoomed out Fig. \ref{fig:featuresA_zoom}.}
    \label{fig:featuresA_big}
\end{figure}

\begin{figure}[!h]
    \centering
    \textbf{BTC/USDT market and its representation - zoomed in.}
    \includegraphics[width=\textwidth]{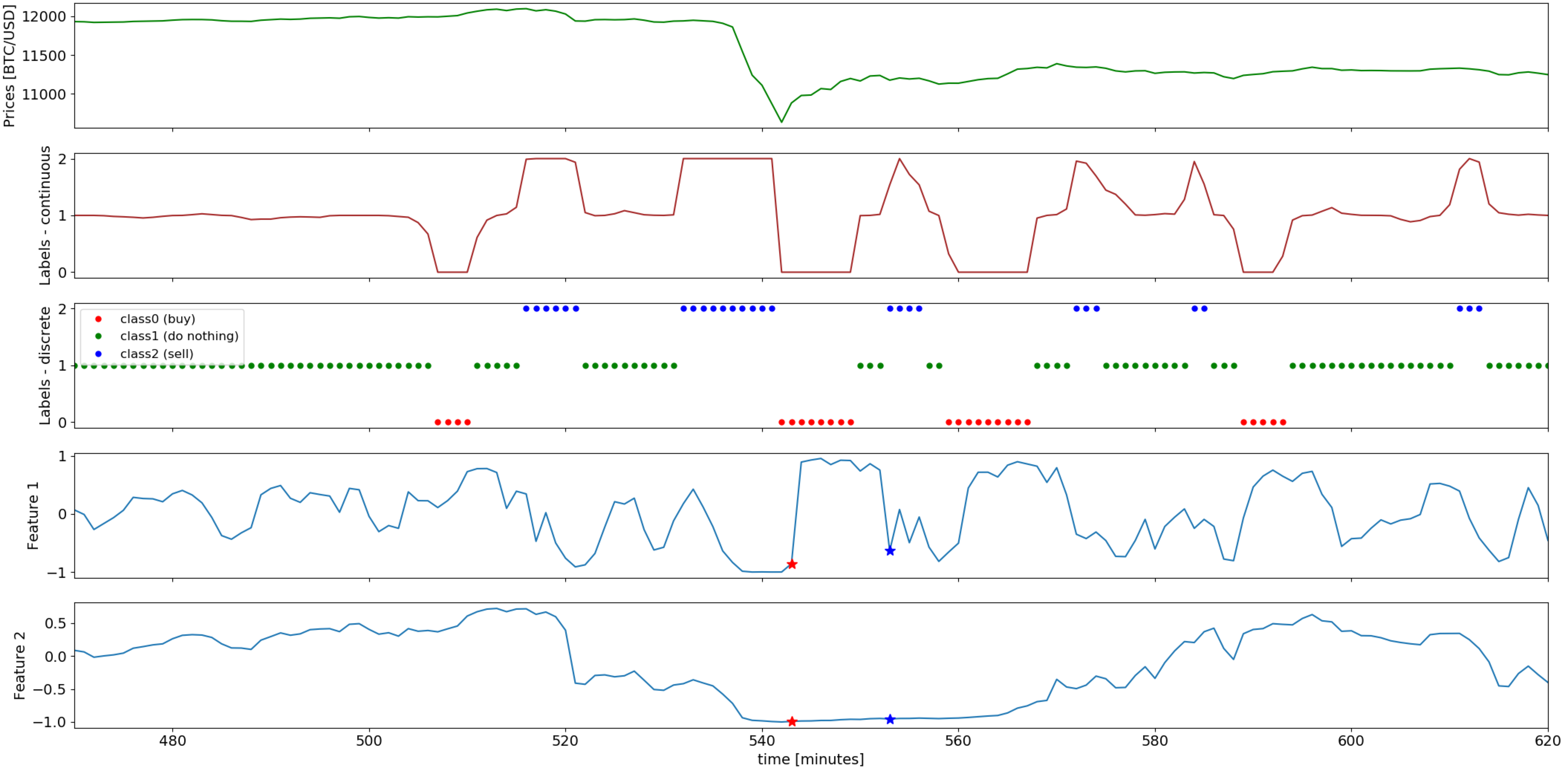}
    \caption{ Fig. \ref{fig:featuresA_big} - zoomed in. The red and the blue star indicate two different market states label-class-wise but with similar feature values in Eq. \ref{eqn:featuresA} market representation. The two features fail to resolve the cross-label-classes which is the central problem of the market representation through feature selection. }
    \label{fig:featuresA_zoom}
\end{figure}

An approximator sees market states only through
their market representation. It is essential that features used in the market representation will differ in values if they encounter different classes of our label of choice. To measure these differences we use the L1 distance between corresponding vectors of features values. The market representations are illustrated on Figures \ref{fig:featuresA_big} and \ref{fig:featuresA_zoom}
\FloatBarrier

\begin{figure}[!h]
    \centering
    \textbf{Cross-label-class distances histogram.}
    \includegraphics[width=\textwidth]{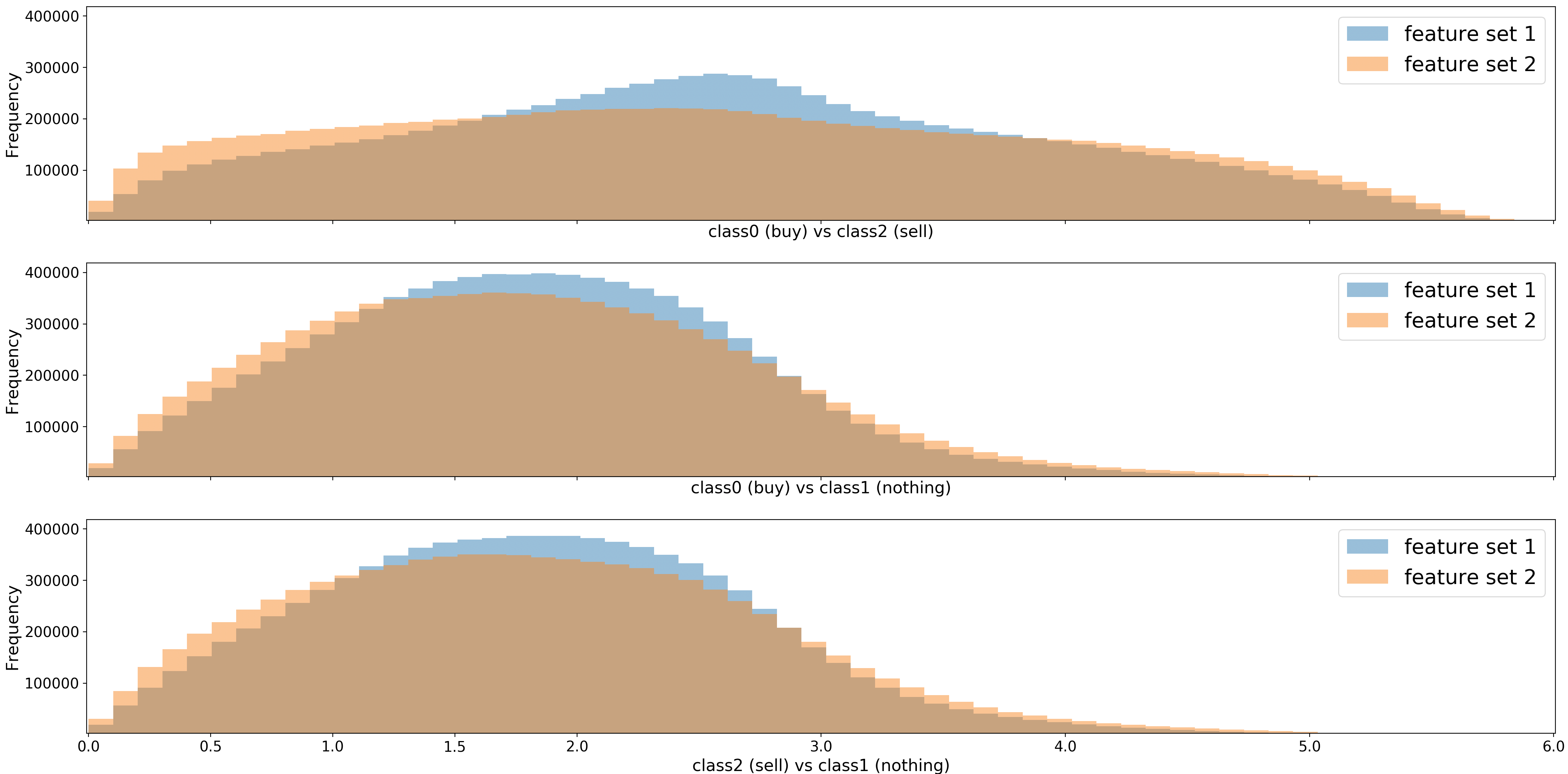}
    \caption{Histogram of distances in Eq. \ref{eqn:featuresA} (feature set 1) and Eq. \ref{eqn:featuresB} (feature set 2) market representations. Both representations contain a lot of cross-label-class pairs with distances close to zero, however, the feature set 1 should be slightly better for an approximator than the feature set 2 because the feature set 1 distances histogram is skewed to the right. 3000 points from each class so 9 million points per histogram}
    \label{fig:hists_features}
\end{figure}

Frequent low values of cross-label-class distances in a given market representation may cause severe problems for an approximator to correctly classify different market states as different label classes.   
Let us look at feature-wise market representation distances across a time period through a histogram of cross-label-class distances - Fig. \ref{fig:hists_features}

\FloatBarrier

Based on Fig. \ref{fig:hists_features} we conclude this subsection saying that the Eq. \ref{eqn:featuresA} market representation is better for the threshold label with 1.2\% price change, 5 minutes time horizon, than the Eq. \ref{eqn:featuresB} market representation. In it important to point out that both representations contain a lot of cross-label-class pairs with distances close to zero, so one should either search for a different feature set from the feature space or change the feature set altogether. 

\subsection{Fixing feature space parameters}
We need a way to quantify goodness of a particular features set to represent a market for a particular label across a time period.
Let us define a following metric for it: \\
\textbf{Label separation power of a feature set}: inverse of an area under a cross-label-class distances histogram (like in Fig. \ref{fig:hists_features}) weighted by a function to only select values relatively close to zero. Choice of the weighting function depends on the label and the numbers of feature in the feature space.

Choosing a particular feature set from a feature space for a given label is done through maximising their label separation power with Bayesian Optimization \cite{bayesian} and HyperBand \cite{hyperband}.

\subsection{Chosen market representation.} \label{subs:chosen_market_representation}
Feature space used in later parts of the paper contains 28 standard price and volume indicators and is formally defined in \nameref{subs:appendix_2}.

Threshold label with 1.2\% price change and 5 minutes time horizon is one of the labels which we used for the analysis. The selected feature set has the following cross-label-class distance histogram:

\begin{figure}[htb]
    \centering
    \textbf{Cross-label-class distances histogram.}
    \includegraphics[width=\textwidth]{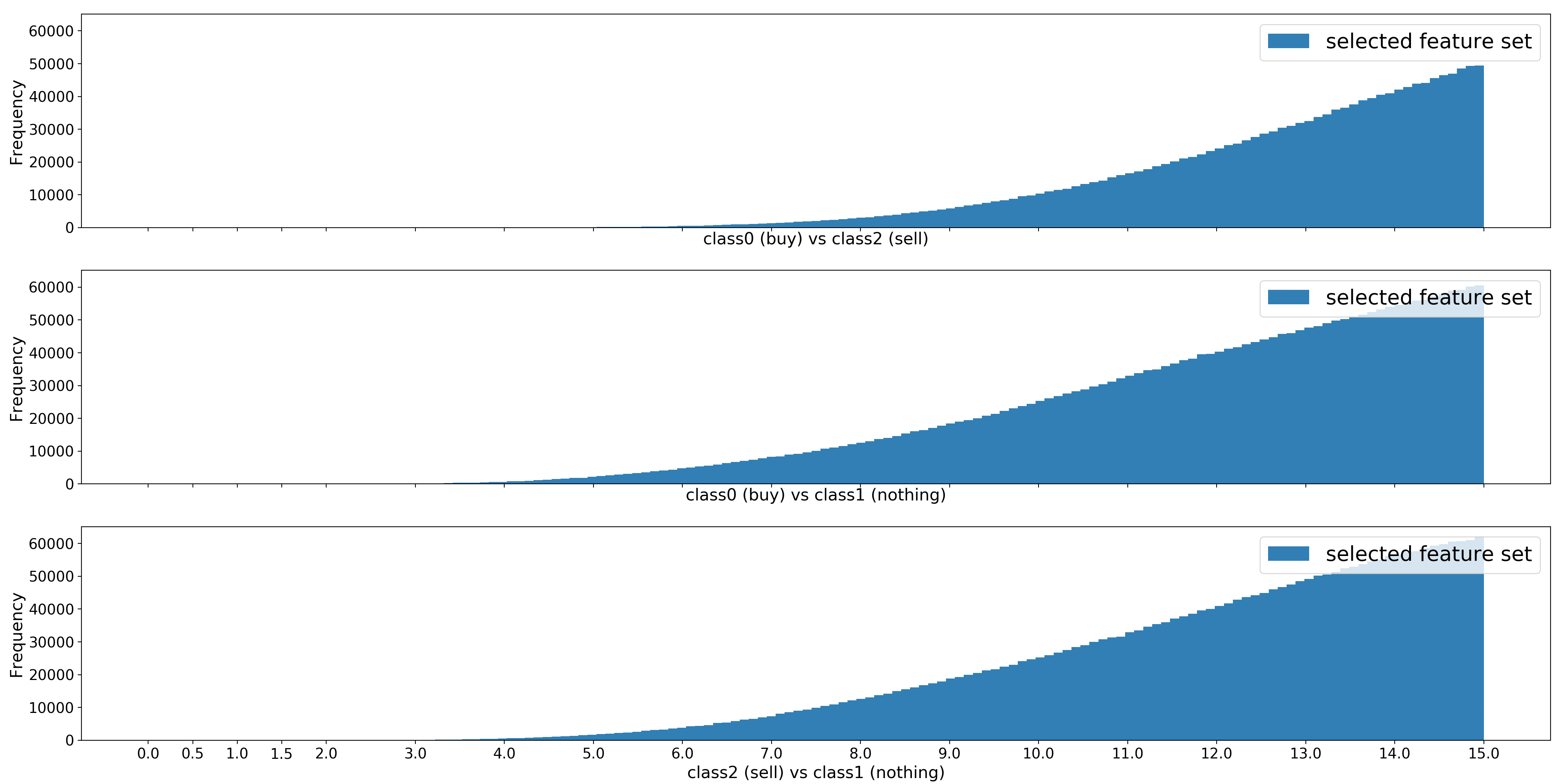}
    \caption{Histogram of distances of the selected features for the threshold label with 1.2\% price change and 5 minutes time horizon. 3000 points from each class so 9 million points per histogram. Note fundamental differences between Fig. \ref{fig:hists_features} and this one. Only \textbf{0.005\%} of cross-label-class distances is below 3 on the same dataset as Fig. \ref{fig:hists_features} histogram.}
    \label{fig:selected_features}
\end{figure}

\begin{figure}[htb]
    \centering
    \textbf{Cross-label-class distances histogram - log scale.} 
    \includegraphics[width=\textwidth]{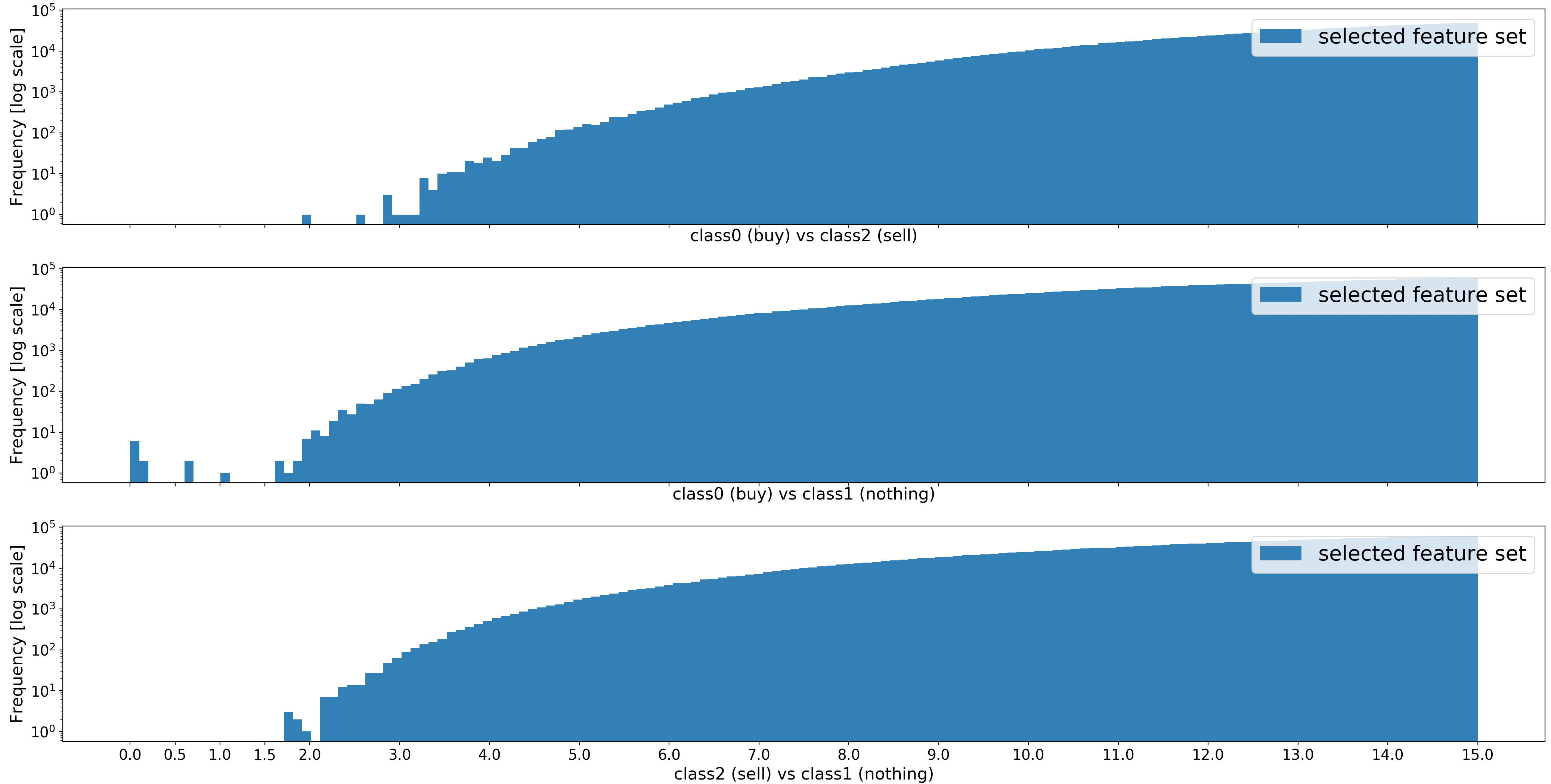}
    \caption{Log scaled Fig. \ref{fig:selected_features} Y-axis-wise. class0 (buy) and class2 (sell) are the easiest to separate - this is in agreement with our intuition.}
    \label{fig:selected_features_log}
\end{figure}

\FloatBarrier

\section{Approximation of labels}
\subsection{Goal of the approximation}
\begin{figure}[htb]
    \begin{center}
    \textbf{Process of live label classification}\par\medskip
    \end{center}
    \includegraphics[width=0.65\textwidth]{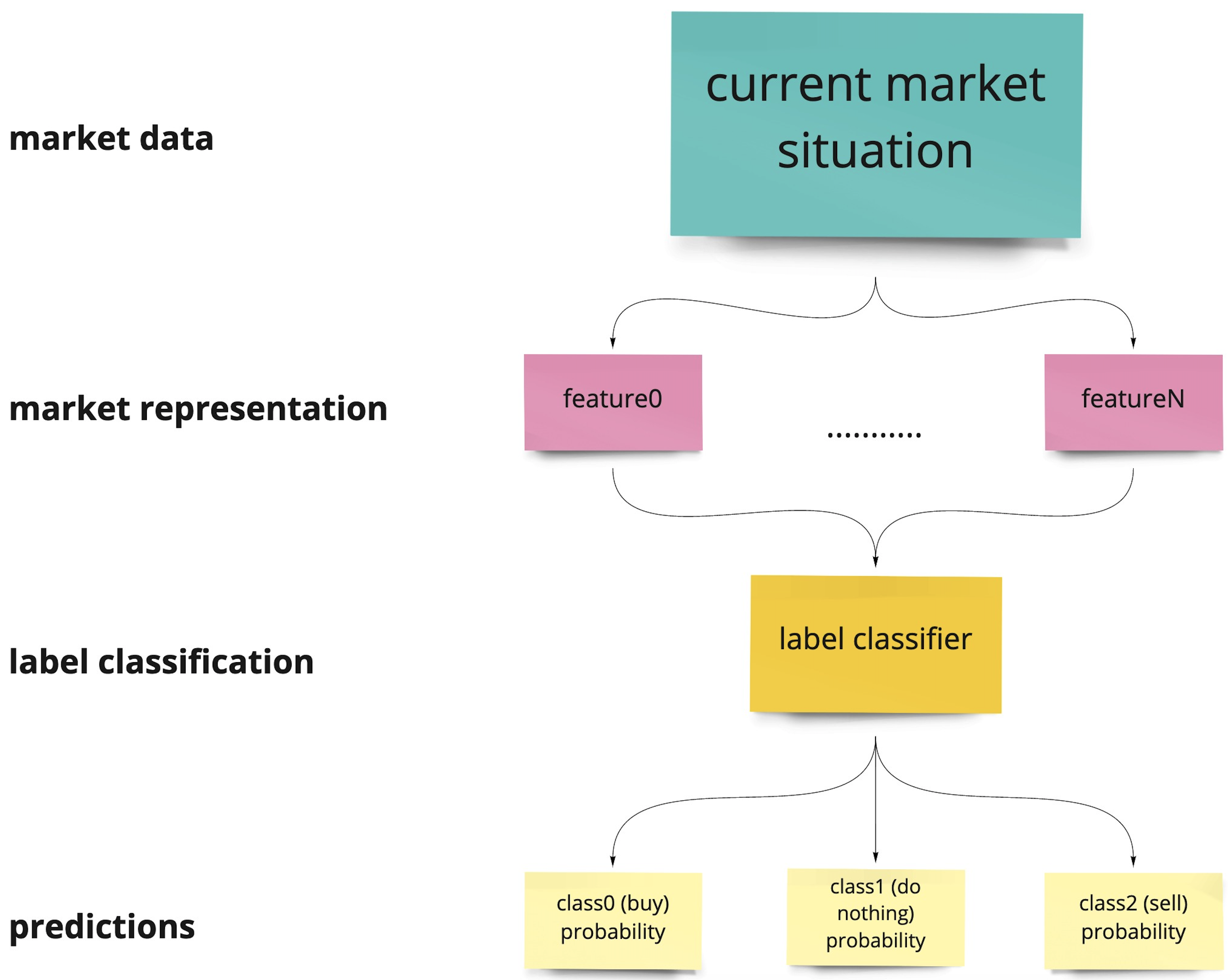}
    \caption{Illustration of label classification using current market data. There is no need to know near future in this process, which makes it possible to perform live.}
    \label{fig:live_ov}
\end{figure}

There are eight labels to approximate (defined in \nameref{subs:appendix_1}) using eight different market representations (from a feature space defined in \nameref{subs:appendix_2}).
Instead of approximating the continuous labels, we choose to classify discrete-label-classes. This way we can focus on identification of the most important three regimes of the labels. In addition, this approach gives us a probabilistic way to determine confidences of our predictions.
Each discrete-label-classes will be assigned a probability of occurring at a given time. The process of training a label classifier using historical market data is illustrated on the Fig.\ \ref{fig:train_ov}, whereas getting the label approximation is illustrated on the Fig.\ \ref{fig:live_ov}:

\begin{figure}[htb]
    \begin{center}
    \textbf{Process of label classifier preparation (training)}\par\medskip
    \end{center}
    \includegraphics[width=0.75\textwidth]{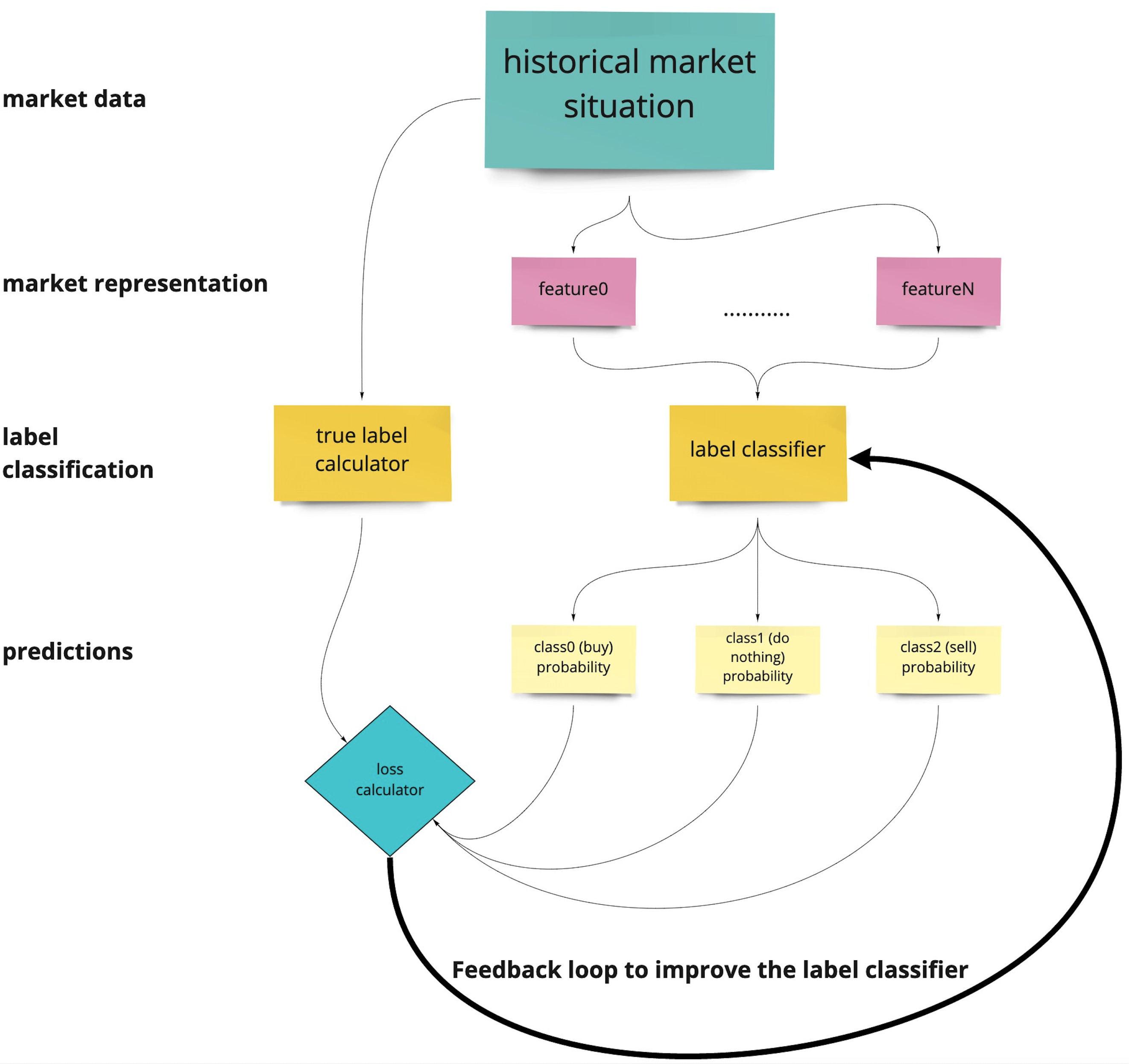}
    \caption{Illustration of training a label classifier using historical market data. Calculating the features requires knowledge of what happened in recent past, however, calculating true labels requires also knowledge of the near future.}
    \label{fig:train_ov}
\end{figure}

\FloatBarrier

\subsection{Chosen algorithm}
 Based on Fig.\ \ref{fig:selected_features} we see that an accurate approximator for this label and this feature set is possible to built but has to be non-linear. We conclude the same for the other seven labels. Because of a high number of datapoints in our training dataset (exact numbers in \nameref{subs:exprimental_setup})  and the requested non-linear behaviour we have decided to use a neural network classifier and a supervised learning algorithm. For hyper-parameter optimisation we used previously mentioned Bayesian Optimization \cite{bayesian} and HyperBand \cite{hyperband}.
 The loss calculator, which appears on Fig.\ \ref{fig:train_ov}, is built based on a concept called \textit{loss scaling} which scales loss based on continuous labels. The central idea is to make class0 (buy) and class2 (sell) prediction accuracy  more significant than class1 (do nothing) in the feedback loop to the label classifier during the training. This is an essential step because of heavy class-unbalance in the labels we have chosen. We construct the scaling in such a way that the sum of loss scale factors associated with class0 (buy) and class2 (sell) is equal to the sum of loss scale factors for class1 (do nothing). In addition, we reduce the loss scaling in-between 0-1 and 1-2 continuous label to make the training focus on clearer buy/do nothing/sell signals \{0,1,2\}.  
 
 \begin{figure}[htb]
    \centering
    \includegraphics[width=\textwidth]{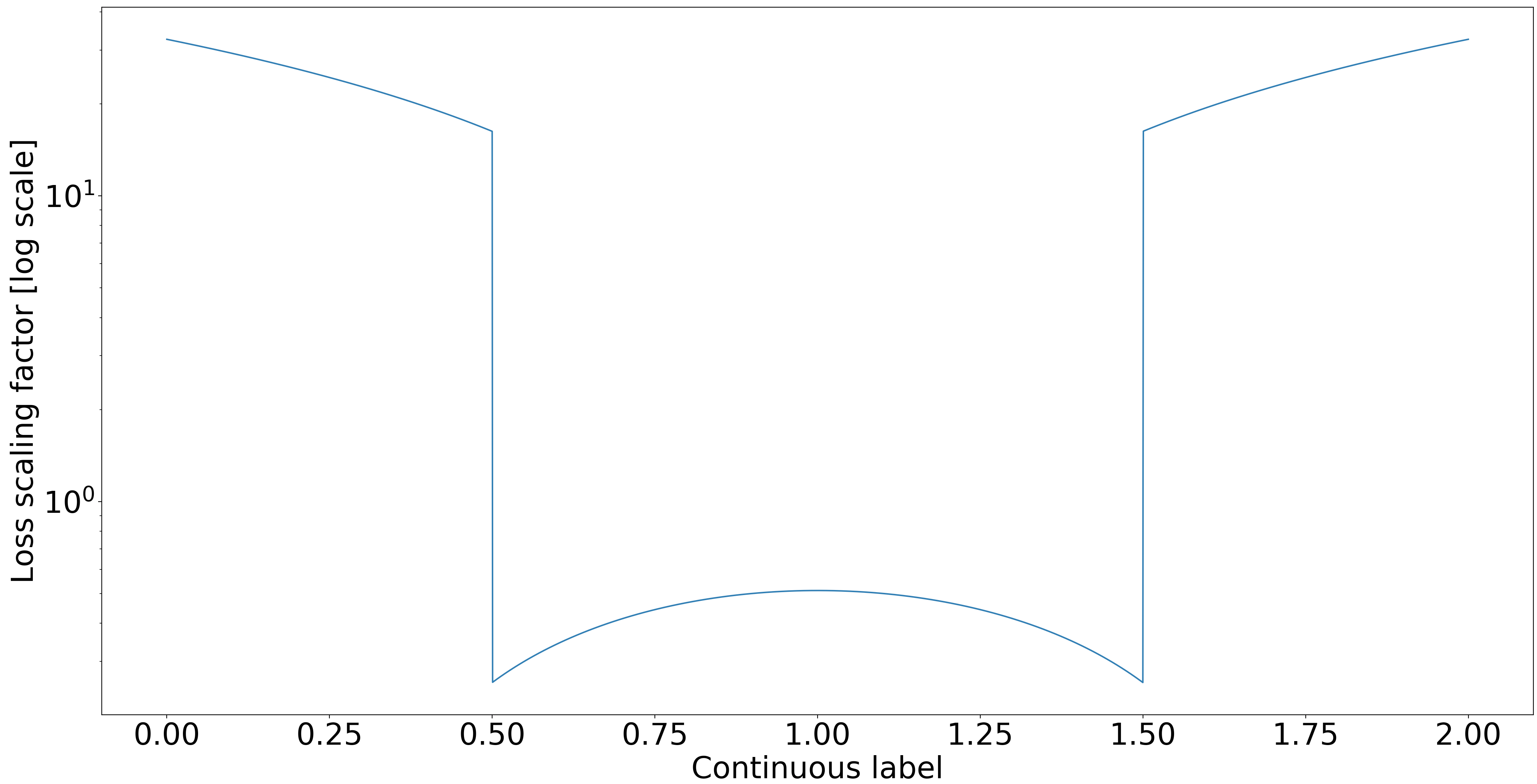}
    \caption{Continuous label based loss scaling factors used during the training process (Fig.\ \ref{fig:train_ov}). The sum of loss scale factors associated with class0 (buy) and class2 (sell) is equal to the sum of loss scale factors for class1 (do nothing). We reduce the loss scaling in-between 0-1 and 1-2 continuous label to make the training focus on clearer buy/do nothing/sell signals (0,1,2). }
    \label{fig:loss_scaler}
\end{figure}
 
\FloatBarrier
 
\subsection{Classifier evaluation}
\begin{figure}[htb]
    \centering
    \textbf{Neural network output}
    \includegraphics[width=\textwidth]{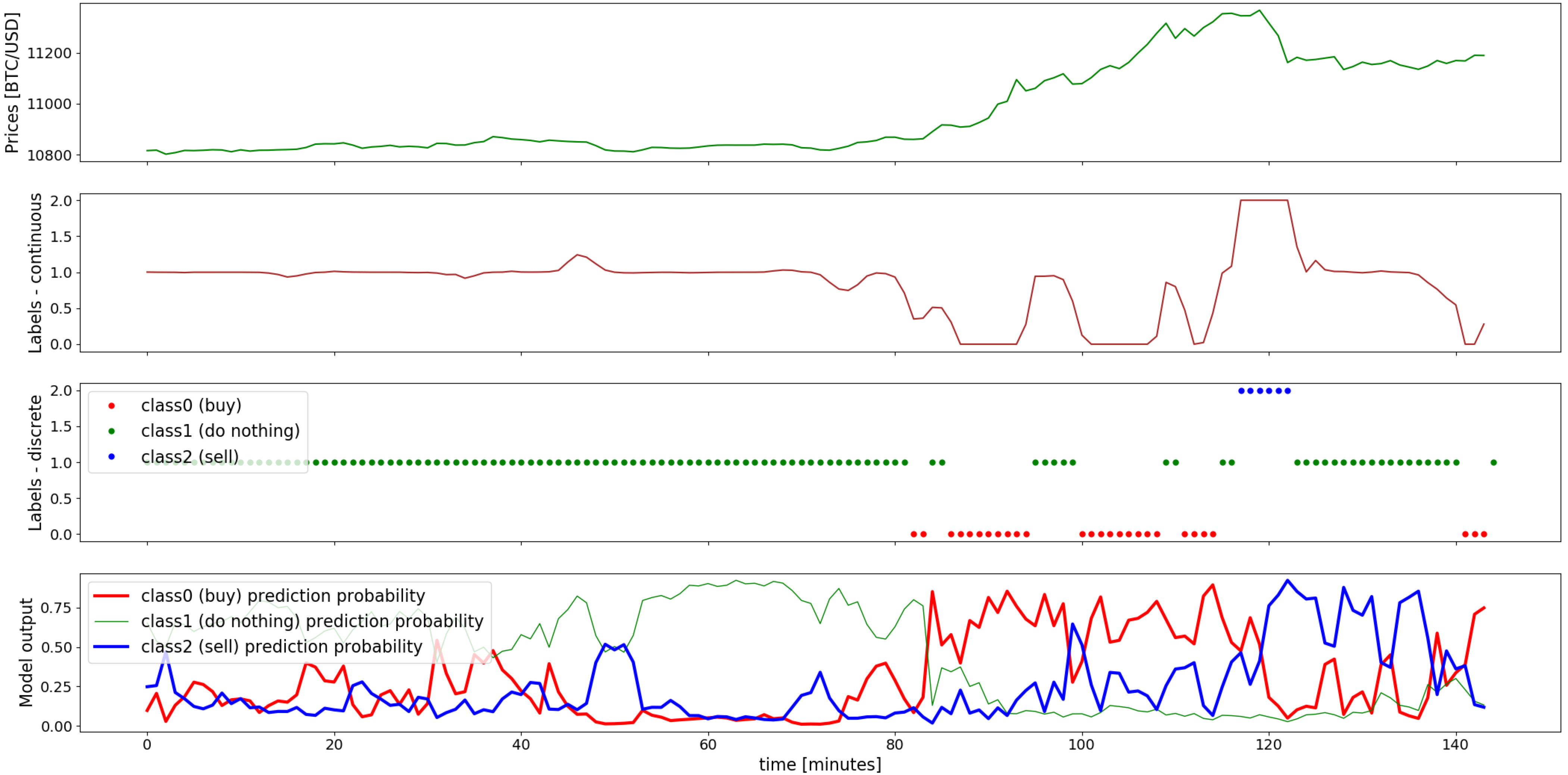}
    \caption{Prices, corresponding true labels and neural network model output predictions. Class0 (buy) in red, class1 (do nothing) in green, and class2(sell) in blue for the threshold label with 1.2\% price change and 5 minutes time horizon. Out-of-sample data fragment. }
    \label{fig:labels_nn}
\end{figure}

A trained classifier acting on unseen data is illustrated on Fig.\ \ref{fig:labels_nn}. Apart from industry-standard metrics like generalisation and confusion matrix coefficients, we also study our classifiers through Shapley Values \cite{prado2020a}\cite{shapley1953a}\cite{Štrumbelj2014}. This approach enables to see impact of a particular feature on the model output. If at this point, the data would not comply with our intuitions, we would not have chosen this particular feature space and the algorithm for label approximation for further experiments.

\FloatBarrier

\section{Trading strategy}
\label{sec:trading_strat}
\begin{figure}[htb]
    \centering
    \includegraphics[width=\textwidth]{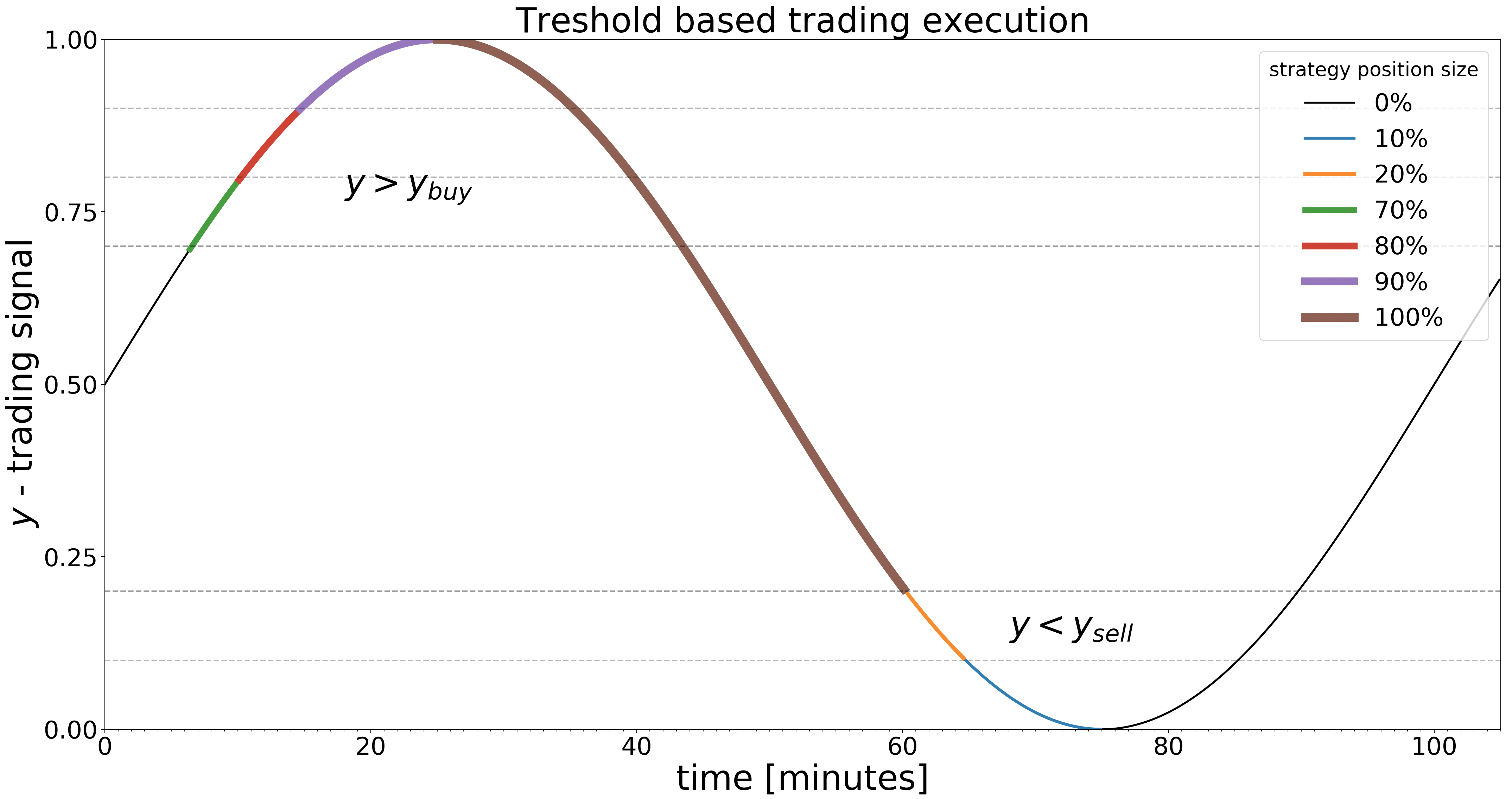}
    \caption{Illustration of the way our trading strategy changes the long position size based on a trading signal $y$. Thresholds for the illustration: $y_{buy}$ = 0.75, $y_{sell}$ = 0.25, $y_{width}$ = 0.10. }
    \label{fig:trading_execution}
\end{figure}

Our strategies are based on linear combinations of 16 model outputs - class0(buy) and class2(sell) of each 8 labels. We map it to $[0,1]$ using a map:

\begin{equation}
        \phi(x)=
        \left\{ \begin{array}{ll}
            (x+1)/2 & |x| < 1 \\
            sgn(x) & \text{otherwise}
        \end{array} \right.
\end{equation}
to acquire a trading signal $y$:

\begin{equation}
    y = \phi(W^T X)
\end{equation}

where $y$ is a trading signal,  $W$ is a column of weights, $X$ is a column of 16 model outputs - class0(buy) and class2(sell) of each 8 labels.
We interpret the output value $y$ as \textbf{a desired long position} on an asset.
To execute trades, we use three thresholds: $y_{buy}$ to buy, $y_{sell}$ to sell and $y_{width}$ to prevent execution of relatively small transactions. 19 constrained parameters in total. The parameter space are formally defined in \nameref{subs:appendix_3}. Fig.\ \ref{fig:trading_execution} illustrates threshold-based trading execution. We increase (decrease) the long position if the trading signal $y$ is above (below) $y_{buy}$ ($y_{sell}$) and a distance between our previous position and the desired position is greater than $y_{width}$.

\FloatBarrier

\section{Experiment}
We ran an experiment of backtesting \nameref{sec:trading_strat} with 20 thousand different weights columns. The goal was to check performances in the \textit{past} dataset and check their generalisation on the \textit{future} dataset. Exact definitions of the used datasets are in the \nameref{subs:exprimental_setup} section. To acquire the weights we ran Bayesian Optimization \cite{bayesian} and HyperBand \cite{hyperband} with a task of producing weights with high performance in the \textit{past} dataset. Because of random nature of the HyperBand algorithm we have acquired a full spectrum of strategies - from bad to good performance-wise.
\subsection{Experimental setup}
\label{subs:exprimental_setup}

We based our experiment on BTC/USDT tick-by-tick transaction data recorded on the Binance exchange.
We divided our dataset into three parts \\ (dd-mm-yyyy):
\begin{itemize}
    \item 01-01-2018 : 31-12-2019: Classifier \textit{train\&evaluate} dataset
    \item 04-01-2020 : 05-05-2020: Strategy \textit{past} dataset
    \item 09-05-2020 : 19-09-2020: Strategy \textit{future} dataset (out-of-sample)
\end{itemize}
Gaps in-between datasets were designed to prevent the look ahead bias. We aggregate the data into 1 minute datapoints.  
Transactions in our backtesting use \textit{the next open price} to execute orders and carry a flat transaction fee equal to \textbf{0.05\%}. This is a realistic estimation of a transaction cost which can be achieved on an exchange.
\\

\subsection{Results - single strategy performances} \label{subs:results_smp}
\begin{figure}[htb]
    \centering
    \includegraphics[width=\textwidth]{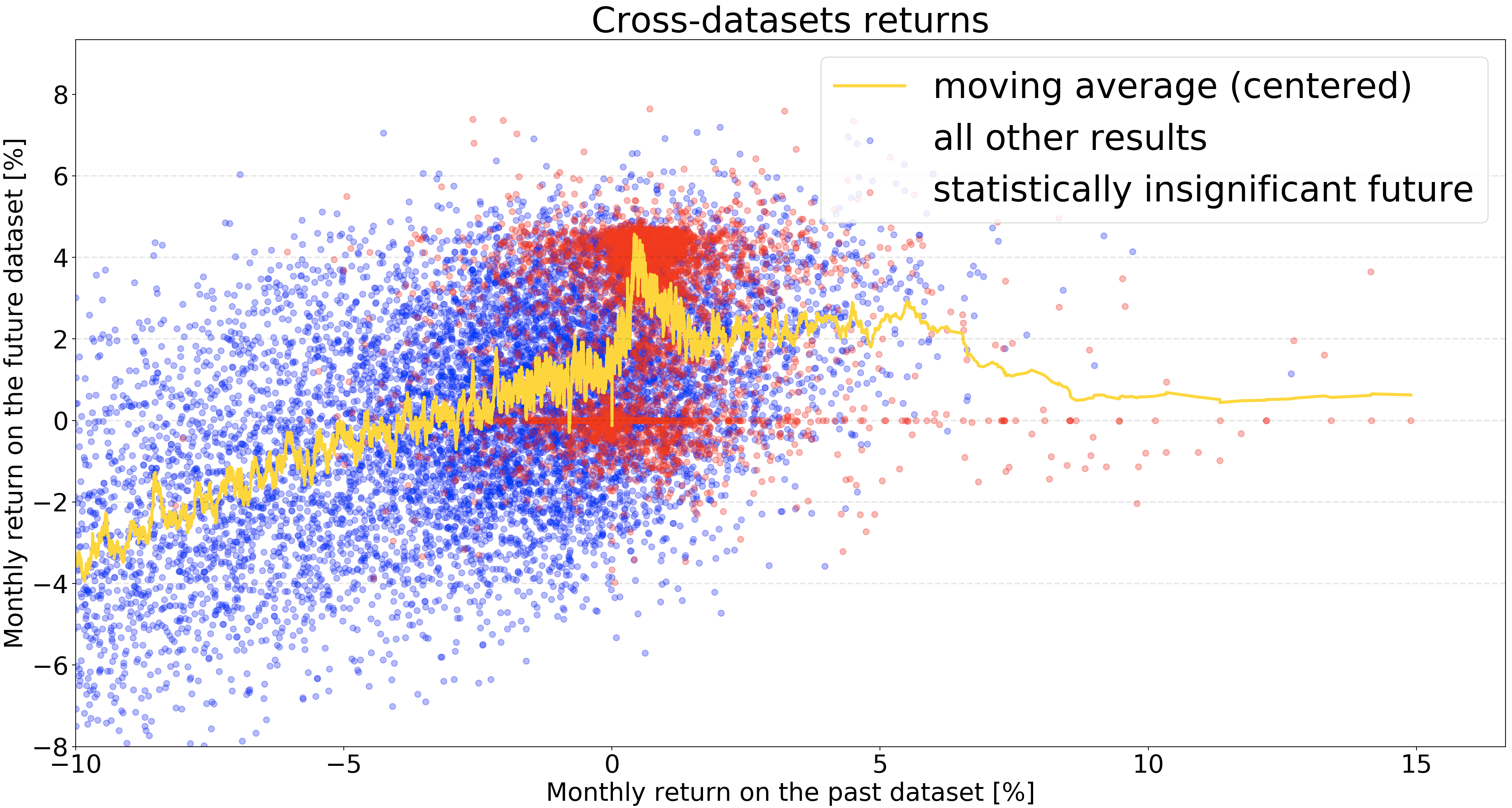}
    \caption{Cross-dataset returns of strategies. Returns on the \textit{past} and \textit{future} datasets are on the X and Y axis respectively. Statistically insignificant future red label corresponds to strategies that resulted in less than 5 trades per month on the \textit{future} dataset}
    \label{fig:cross_returns_transactions}
\end{figure}

We have to find a selection metric for our strategies.
First, let us talk about cross-datasets returns, where we put \textit{past} dataset strategy returns against corresponding \textit{future} dataset strategy returns. The comparisons are illustrated on the Fig.\ \ref{fig:cross_returns_transactions}.

\FloatBarrier
As illustrated on Fig.\ \ref{fig:cross_returns_transactions}, a \textit{past} dataset return cannot be used as a metric to select strategies with promising results in the future. The positive correlation brakes down around 5\% montly return on the \textit{past} dataset and the region with highest \textit{future} returns is filled with statistically insignificant \textit{future} performances. We need a better strategy selection method.

Let us introduce \textit{score function} as follows:
\begin{equation}
    S = MP - TP - MCIP
\end{equation}
where:
\begin{itemize}
    \item S: strategy score
    \item MR: montly return
    \item TP: transaction penalty - if number of transaction per month on the \textit{past} dataset is lower than 30, then it is equal to the missing transactions per month
    \item MCIP: mean capital involvement penalty - if a mean capital involvement (mean long position) is lower than $25$\%, then it is equal to a half of missing percentages
\end{itemize}
The goal of the \textit{score function} is to map the problematic low-return or statistically insignificant regions (illustrated on the Fig.\ \ref{fig:cross_returns_transactions}) to low scores but preserving the positive  performance correlation structure. Parameter values of the \textit{score} function were chosen intuitively, before the cross-datasets studies. The number 30 in the transaction penalty was chosen simply because for Bitcoin it is the number of trading days in a month, and all of the used features can drastically change intra-day.  An idea how to modify the score function, to make it less accidental is presented in \nameref{sec:pb_modifications}.

Cross-score performances are illustrated on the Fig.\ \ref{fig:score_vs_return} down below:

\begin{figure}[htb]
    \centering
    \includegraphics[width=\textwidth]{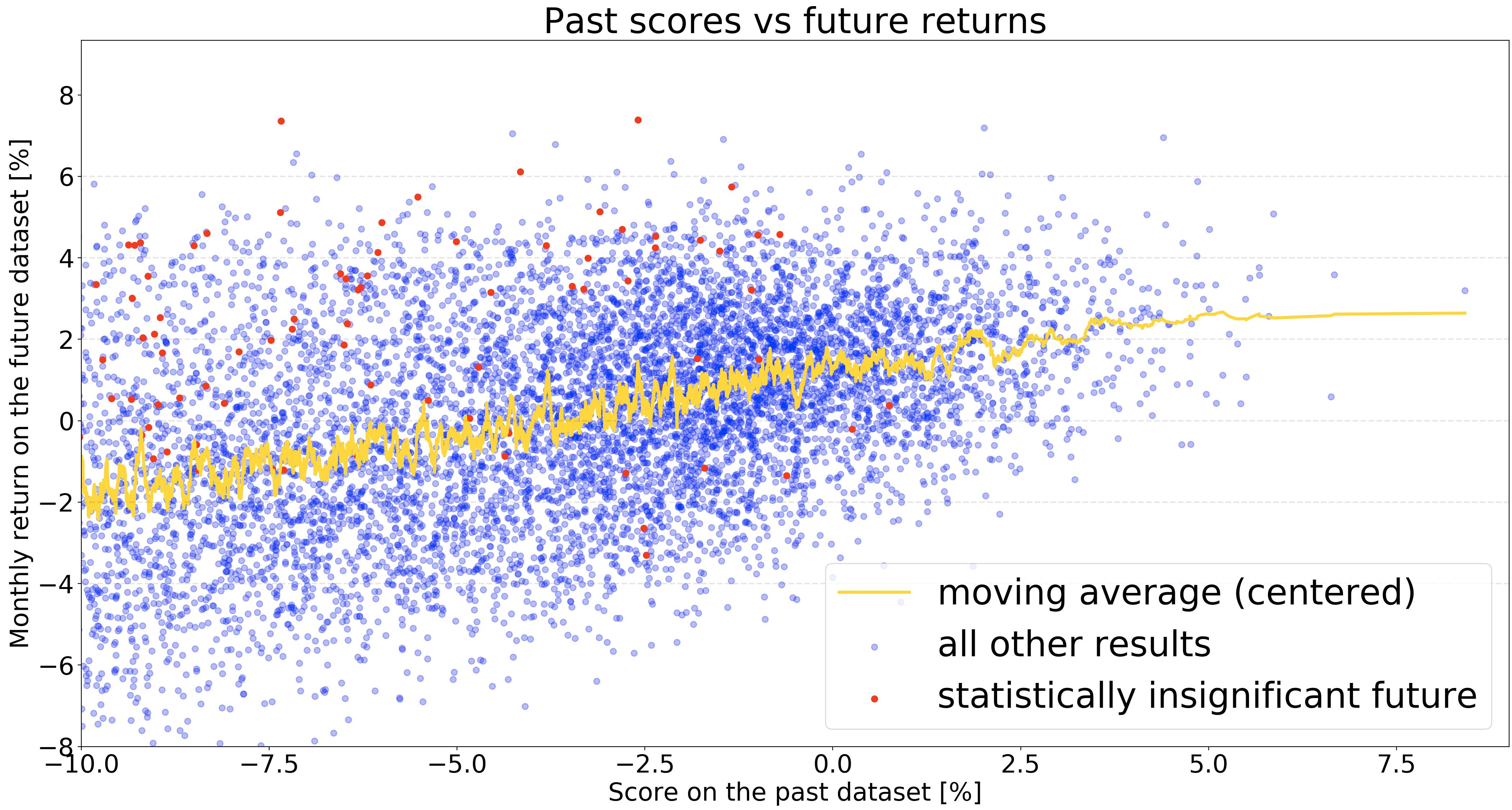}
    \caption{Cross-dataset performances of strategies. Scores on the \textit{past} dataset and returns on the \textit{future} dataset are on the X and Y axis respectively. Statistically insignificant future red label corresponds to strategies that resulted in less than 5 trades per month on the \textit{future} dataset.}
    \label{fig:score_vs_return}
\end{figure}
\FloatBarrier
The higher the \textit{past} dataset score the higher (on average) the \textit{future} dataset monthly returns. The statistically insignificant regions were mapped out of the illustrated on Fig.\ \ref{fig:score_vs_return} region. The \textit{score} can now be used as a selection metric for our trading strategies.

\subsection{Results - strategy ensemble performances }

\begin{figure}[htb]
    \centering
    \includegraphics[width=\textwidth]{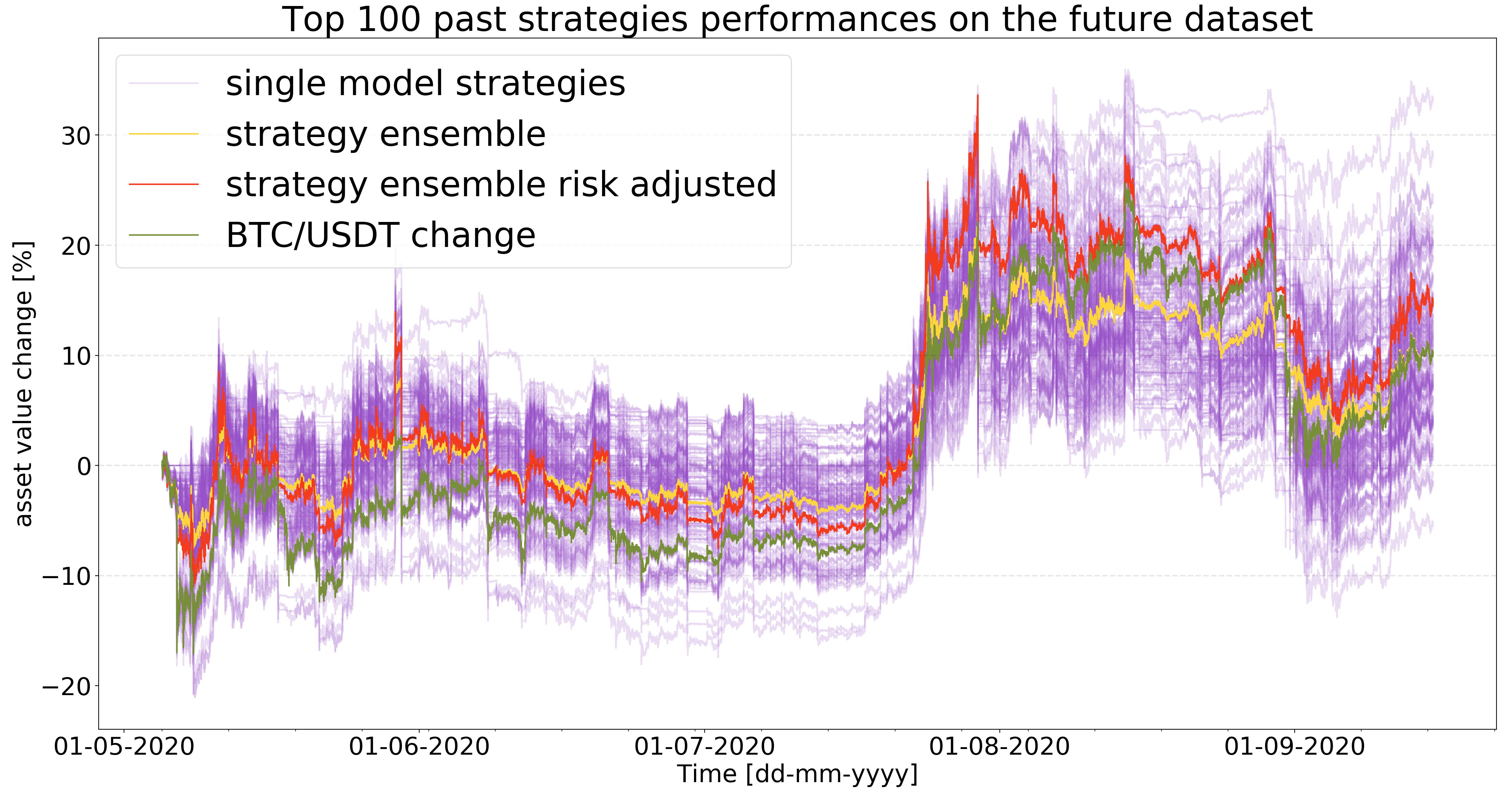}
    \caption{Performance of top 100 score-wise strategies and a corresponding strategy ensemble on the \textit{future} dataset. Risk adjustment is performed through scaling the strategy ensemble returns by  $\frac{\sigma _{BTC/USDT}}{\sigma _{ensemble}}$ where $\sigma$ is the standard deviation of returns. }
    \label{fig:top100}
\end{figure}

\begin{figure}[htb]
    \centering
    \includegraphics[width=\textwidth]{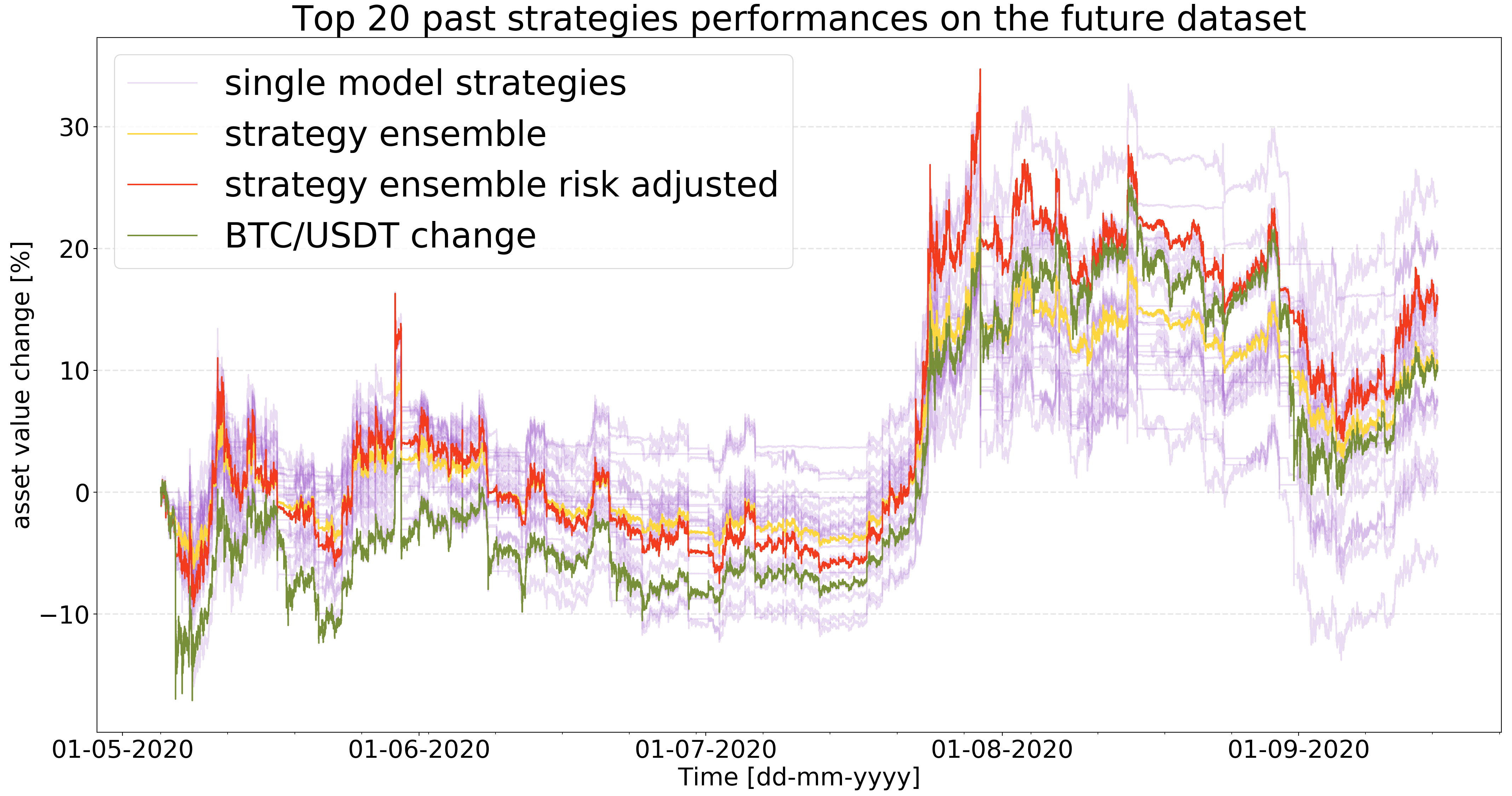}
    \caption{Performance of top 20 score-wise strategies and a corresponding strategy ensemble on the \textit{future} dataset. Risk adjustment is performed through scaling the strategy ensemble returns by  $\frac{\sigma _{BTC/USDT}}{\sigma _{ensemble}}$ where $\sigma$ is the standard deviation of returns. }
    \label{fig:top20}
\end{figure}

\begin{figure}[htb]
    \centering
    \includegraphics[width=\textwidth]{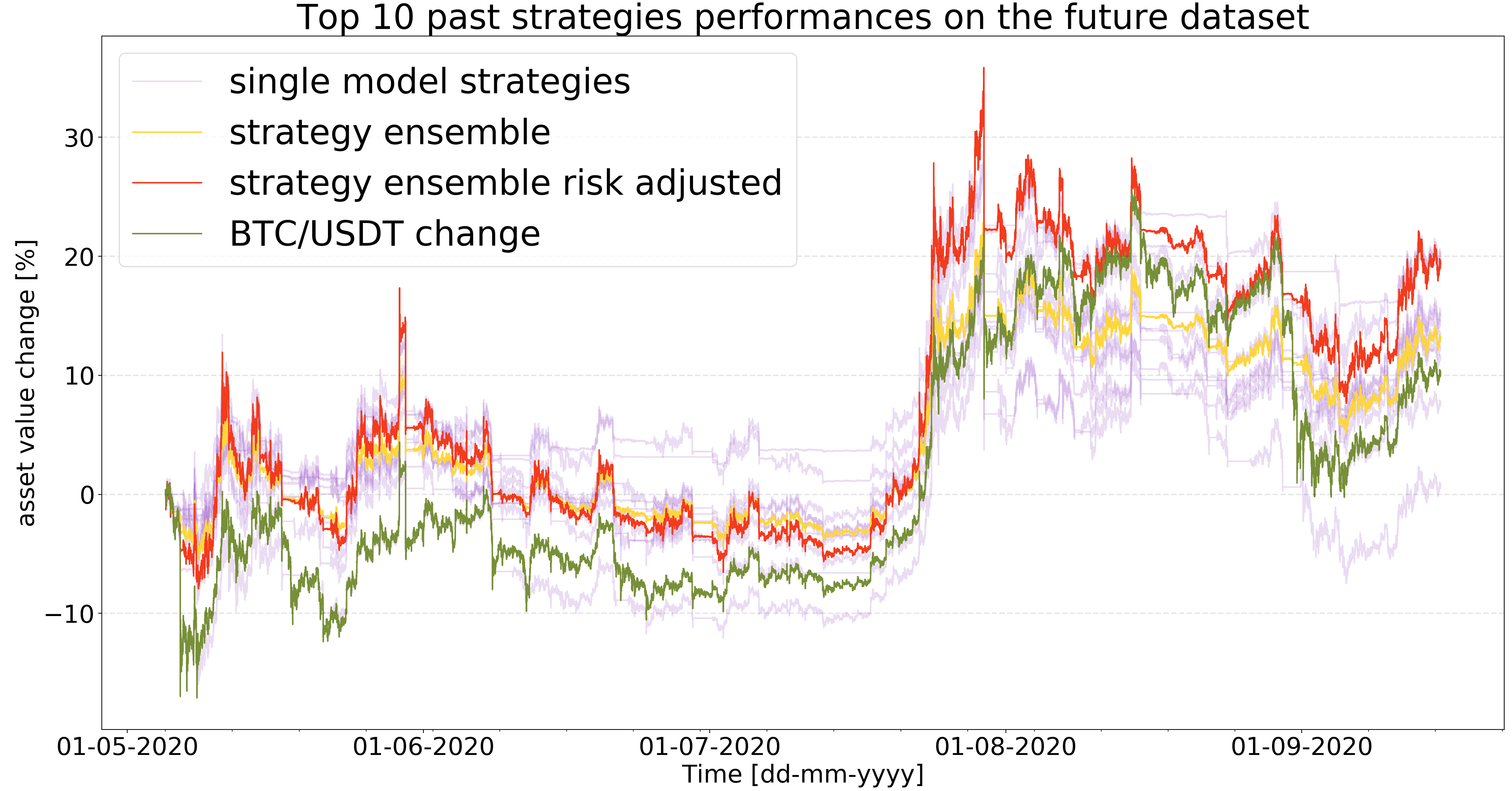}
    \caption{Performance of top 10 score-wise strategies and a corresponding strategy ensemble on the \textit{future} dataset. Risk adjustment is performed through scaling the strategy ensemble returns by  $\frac{\sigma _{BTC/USDT}}{\sigma _{ensemble}}$ where $\sigma$ is the standard deviation of returns. }
    \label{fig:top10}
\end{figure}

\begin{figure}[htb]
    \centering
    \includegraphics[width=\textwidth]{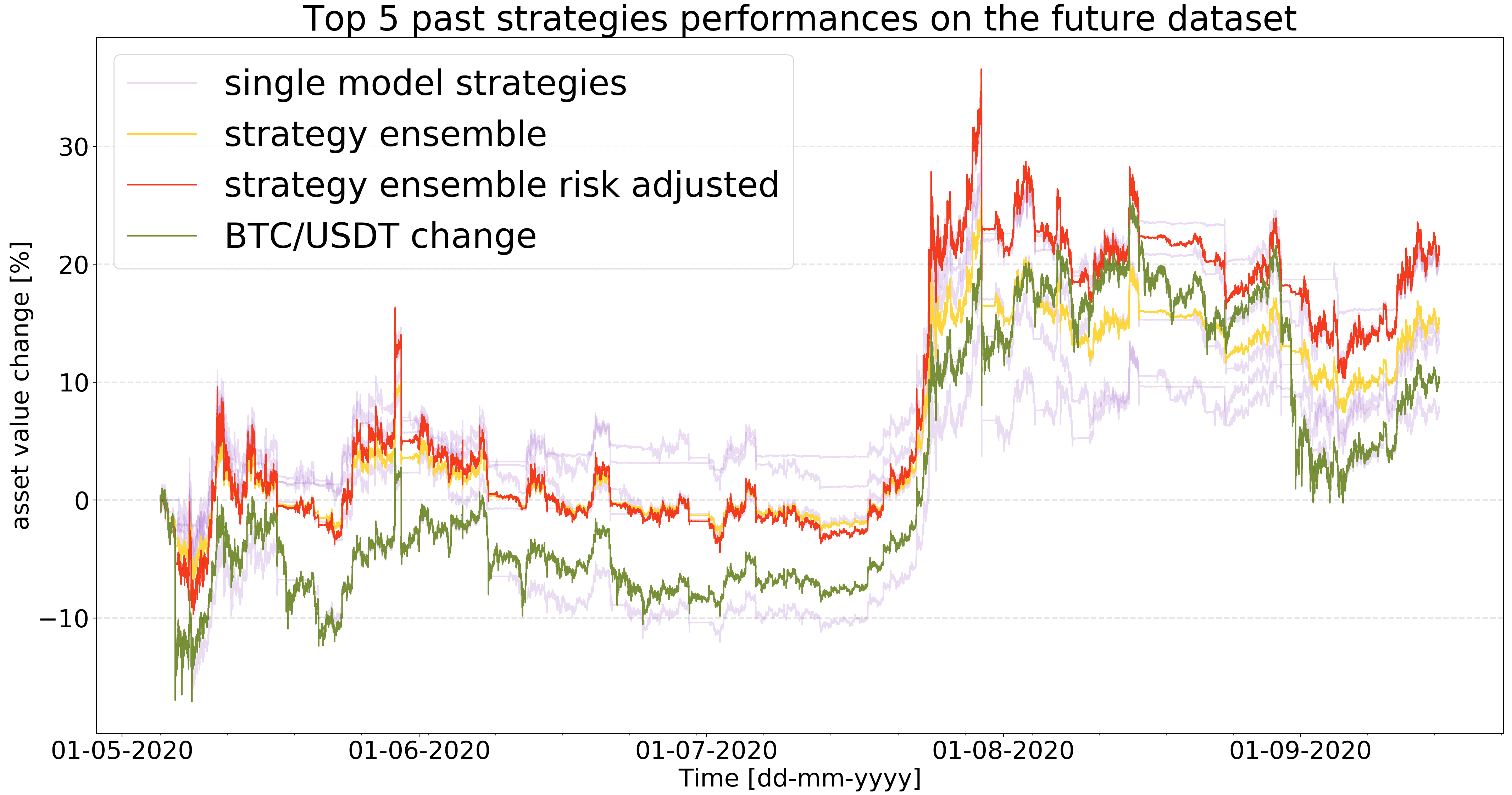}
    \caption{Performance of top 5 score-wise strategies and a corresponding strategy ensemble on the \textit{future} dataset. Risk adjustment is performed through scaling the strategy ensemble returns by  $\frac{\sigma _{BTC/USDT}}{\sigma _{ensemble}}$ where $\sigma$ is the standard deviation of returns. }
    \label{fig:top5}
\end{figure}

Strategy ensemble backtests using Top100, Top20, Top10, and Top5 \textit{past} score-performance-wise models are illustrated on Fig.\ \ref{fig:top100}, Fig.\ \ref{fig:top20}, Fig.\ \ref{fig:top10}, and Fig.\ \ref{fig:top5} respectively.

\FloatBarrier
Both the returns and risk-adjusted returns are increasing as the average \textit{past} score-performance increases.

\section{Conclusions}
Performances of created strategies increase in terms of return and risk-adjusted return on the out-of-sample \textit{future} dataset as the \textit{past} \textit{score}-performance increases. Using top \textit{score}-performance-wise strategies we achieved exceptional market-beating results. As of right now, using a framework which we have described can lead to further improvements of capital allocations of institutional investors with access to market data and computational power. 

\section{Future work}
\label{sec:pb_modifications}
\textbf{Making transaction rates in the \textit{score function} definition dataset dependent}, because static rates lead to ruling out potentially high performance strategies if they do not comply with dataset's dynamics. The optimal transaction rate should be based on characteristics of the \textit{past} dataset - i.e.\@ average volatility - and in general should not be hard-coded.

\textbf{Changing strategies \textit{on the fly}} should further increase performance by swapping under-performing single model strategies with more promising substitutes. In this approach, the strategies are ranked for selection based on their up-to-date past performances. The effective \textit{past} dataset would change periodically.

\textbf{More sophisticated \textit{feature space}} would potentially lead to better classifiers and enable detection of sub-minute movements. 

\textbf{More sophisticated trading strategies} - our linear combination was selected to reduce complexity. We now look into more complex solutions which are still relatively easy to interpret.

\textbf{Running computations for longer} to find higher \textit{past} score-performance-wise strategies should further increase out-of-sample performances. 

\section*{Acknowledgement}
We would like to thank David Klemm for his support, discussions, and the access to his computational grid. 

\appendix
\section{Appendix: selected labels}\label{subs:appendix_1}
\textbf{The 8 chosen labels} can be categorised into 2 subcategories: threshold labels (see \nameref{sec:labeling}) and local extrema labels.

\textbf{Threshold labels short descriptions}:
\begin{itemize}
    \item 1.2\% price change in the next 5 minutes
    \item 1.2\% price change in the next 60 minutes
    \item 2.2\% price change in the next 2 minutes
    \item 3\% price change in the next 5 minutes
    \item 3\% price change in the next 60 minutes
\end{itemize}
The remaining 3 local extrema labels are a custom construct and are a material for a separate paper. Visualise and explore them all through our repository:  \href{ https://github.com/m1balcerak/labels}{GitHub Labels}

\section{Appendix: \textit{feature space}}\label{subs:appendix_2}
The \textit{feature space} (see \nameref{sec:market_representation})  we use consists of 28 functions and is illustrated on the Fig. \ref{fig:feature_space} below:

\begin{figure}[htb]
    \centering
    \textbf{Definition of our \textit{feature space} .}\par\medskip
    \includegraphics[width=\textwidth]{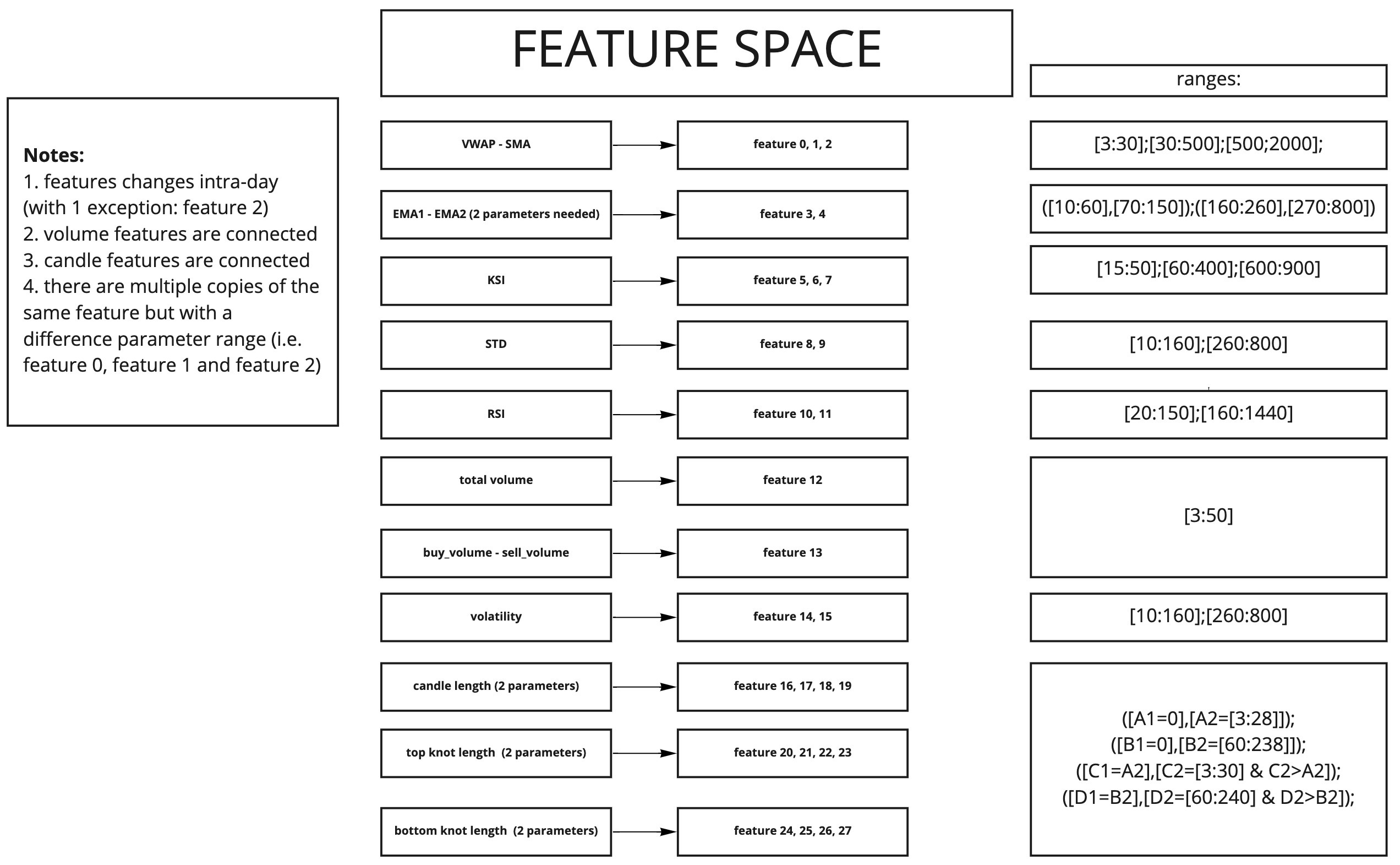}
    \caption{Definition of each of the 28 features and their corresponding parameter ranges that were used during our search for optimal market representations in \nameref{sec:market_representation}. Parameters are integer only and represent minutes. }
    \label{fig:feature_space}
\end{figure}
Ranges of possible parameters and types of indicators are based on our domain knowledge.

\FloatBarrier
\section{Appendix: \textit{strategy space}}\label{subs:appendix_3}
The \textit{strategy space} (see \nameref{sec:trading_strat}) consists of 19 parameters: 3 thresholds and 16 weights. \\ \textbf{Strategy thresholds:} $y_{buy} \in [0.7,1]$, $y_{sell} \in [0,0.3]$, $y_{width} \in [0.02,0.1]$. \\ \textbf{Weights}: $w_n \in [-1,1]$.

\printbibliography
\end{document}